\documentclass[onecollarge,natbib]{svjour2}
\bibpunct{[}{]}{,}{n}{}{,} 
\smartqed  
\usepackage{graphicx}
\newcommand{\slt}{\!\!\!/}
\newcommand{\sld}{\!\!/}
\newcommand{\beq}{\begin{equation}}
\newcommand{\eeq}{\end{equation}}
\newcommand{\beqa}{\begin{eqnarray}}
\newcommand{\eeqa}{\end{eqnarray}}

\journalname{Few-Body Systems (APFB2011)}
\begin{document}
\title{\boldmath
Role of the $K_1$ meson in $K^0$ photoproduction off the deuteron
}
\author{A. Salam         \and
        T. Mart						\and
        K. Miyagawa
}
\institute{A. Salam \at
              Departemen Fisika, FMIPA, Universitas Indonesia, Depok 16424,
              Indonesia\\ 
              \email{agussalam@fisika.ui.ac.id} 
              \and
           T. Mart \at
              Departemen Fisika, FMIPA, Universitas Indonesia, Depok 16424,
              Indonesia 
              \and
           K. Miyagawa \at
              Simulation Science Center, Okayama University of Science, 1-1
              Ridai-cho, Okayama 700-0005, Japan 
}
\date{Received: date / Accepted: date}
%
\maketitle
\begin{abstract}
Neutral kaon photoproduction off the nucleon and deuteron has been reinvestigated by utilizing the new experimental data on both targets. An isobar model for elementary operator and impulse approximation for the reaction on the deuteron  have been used. The available free parameters in the elementary model have been extracted from both elementary and deuteron data. In contrast to the elementary reaction, fitting the deuteron data requires an inclusion of weighting factor. The result indicates that the angular distribution of the elementary $K^0\Lambda$ process does not show backward peaking behavior.  
\end{abstract}
%
\section{Introduction}
\label{introduction}
Kaon photoproduction on light nuclei provides invaluable information
on the strangeness dynamics in the realm of elementary particles and nuclear matters. Reactions on the
proton, i.e., $p(\gamma,K^+)\Lambda$, $p(\gamma,K^+)\Sigma^0$, and
$p(\gamma,K^0)\Sigma^+$, have been intensively investigated~\cite{MaB00,Mar11,ByT11}, since experimental data are available mostly in these channels~\cite{Boc94,Tra98,Goers:1999sw}. Since there is no free neutron target available, 
one can use deuteron as an effective neutron target. By choosing deuteron as a target, some rescattering effects might appear in the final state of the reaction~\cite{YaM99,SaA04}. These effects introduce some complications in the reaction. Nevertheless, one can look for some appropriate kinematical regions to reduce these effects~\cite{SaM06}.  

Recently, an experiment of neutral kaon photoproduction on the deuterium
has been performed at the Laboratory of Nuclear Science (LNS), in
Sendai~\cite{Tsu08}. In this experiment the cross section of the inclusive process $d(\gamma,K^0)YN$ at photon energies around reaction threshold with forward kaon angles have been measured. In the previous work~\cite{SaM09}, comparison between experimental data and theoretical calculation by using elementary operator of KAON-MAID \cite{MaB00} have been made and it is  shown that the model overpredicts the data, especially in the $n(\gamma,K^0)\Lambda$ channel. A fair agreement with the data could be achieved when the coupling of $\gamma K K_1$ vertex in that channel were set equal to zero. However, such treatment might disturb the cross section in the other channels, e.g., the $p(\gamma,K^0)\Sigma^+$ channel, which also contains the $\gamma K K_1$ vertex. It is the purpose of this paper to reinvestigate kaon photoproduction on the nucleon and deuteron by fitting the model to the new experimental data on the elementary reaction only, as well as simultaneously to the deuteron data from LNS.     

This paper is organized as follows. The formalism for calculating the
transition matrix and observables are shown in Sect.~\ref{formulation}. The
results and comparisons with experimental data are presented in
Sect.~\ref{result}. We close this paper with a conclusion in
Sect.~\ref{conclusion}. 

\section{Formulation}
\label{formulation}

In this work we use the isobar model for the elementary operator which includes 
some important resonance terms besides the Born terms. While this leads to violation of unitarity, this kind of isobar model provides a simple tool to parametrize kaon photoproduction on the nucleon because it is relatively easy to calculate and to use for production on nuclei. In this approach, the photoproduction amplitude can be written as
\begin{eqnarray}
\langle\vec p_{Y}\mu_{Y}
\vert t^{\gamma K}_{\lambda}\vert
\vec p_{N}\mu_{N}\rangle
&=& 
\bar{u}_{\mu_{Y}}
\Big(\sum_{i=1}^{4} A_{i} \Gamma^{i}_{\lambda}\Big)
u_{\mu_{N}}\,,~~
\label{eq-gnky-amplitude}
\end{eqnarray}
where $A_{i}$'s are invariant amplitudes as functions of the Mandelstam variables only. The hyperon and nucleon Dirac spinors are denoted by $u_{\mu_{Y}}$ and $u_{\mu_{N}}$, respectively. The invariant Dirac operators $\Gamma^{i}_{\lambda}$, which are given by
\begin{eqnarray}
\Gamma^{1}_{\lambda} 
& = & {\textstyle \frac{1}{2}} \gamma_{5} 
\left({\epsilon\sld}\!_{\lambda} k \slt 
- k \slt {\epsilon\sld}\!_{\lambda} \right)\,, \\
\Gamma^{2}_{\lambda} 
& = & \gamma_{5} \left[ (2q-k) \cdot \epsilon_{\lambda} P \cdot k
- (2q-k) \cdot k P \cdot \epsilon_{\lambda} \right]\,, \\
\Gamma^{3}_{\lambda} 
& = & \gamma_{5} \left( q\cdot k {\epsilon\sld}\!_{\lambda}
- q \cdot \epsilon_{\lambda} k \slt \right)\,, \\
\Gamma^{4}_{\lambda} 
& = & i \epsilon_{\mu \nu \rho \sigma} \gamma^{\mu} q^{\nu}
\epsilon^{\rho}_{\lambda} k^{\sigma}\,,
\label{eq-gnky-Gamma-matrix}
\end{eqnarray}
are gauge Lorentz invariant pseudoscalars and given in terms of the usual $\gamma$-matrices, the photon momentum $k$, and its polarization vector $\epsilon_{\lambda}$. Here $\lambda$ labels the polarization states, $q$ the meson momentum, and $P=(p'+p)/2$, where $p$ and $p'$ denote initial and final baryon momenta, respectively. The rather lengthy expression of $A_{i}$ can be found in Ref.~\cite{Mar96}.
%
\begin{figure*}
\centering
  \includegraphics[width=0.75\textwidth]{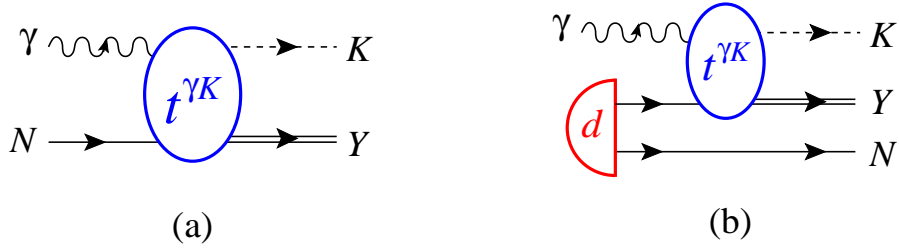}
\caption{Feynman diagrams for kaon photoproduction off a nucleon (a) and a deuteron according to the impulse approximation (b).}
\label{reaction}    
\end{figure*}

We apply the impulse approximation for kaon photoproduction on the deuteron (Fig.~\ref{reaction}). Through out the paper we work in the deuteron rest frame. For the inclusive process $d(\gamma,K^0)YN$ the cross section 
is given by
\begin{eqnarray}
\frac{d\sigma}{dp_{K}d\Omega_{K}} &=&  
\int d\Omega^{\,\rm cm}_{Y}\, 
\frac{m_{Y}m_{N}\vert\vec p_{K}\vert^2\vert\vec p^{\,\rm cm}_{Y}\vert}
{4(2\pi)^2E_{\gamma}E_{K}W}\,
\nonumber\\ &&\times\,
\frac{1}{6} \sum_{\mu_{Y}\mu_{N}\mu_{d}\lambda} 
\big\vert\sqrt{2}
\langle\vec p_{Y}\vec p_{N}\mu_{Y}\mu_{N}
\vert t^{\gamma K}_\lambda\vert
\Psi_{\mu_{d}}\rangle\,
\big\vert^2\,,
\label{eq-gdkyn-inclusive-cross-section}
\end{eqnarray}
where $W^2=(P_{d}+Q)^2$ with $Q=p_{\gamma}-p_K$ is the momentum transfer and $\vert\vec p^{\,\rm cm}_{Y}\vert$ 
is the hyperon momentum calculated in the center of mass frame of the two
final baryons. For the radial part of the deuteron wave function $\Psi_{\mu_d}$ we use
the Nijmegen models~\cite{RiS99}. 

\section{Result}
\label{result}

\begin{figure*}
\centering
  \includegraphics[width=0.70\textwidth]{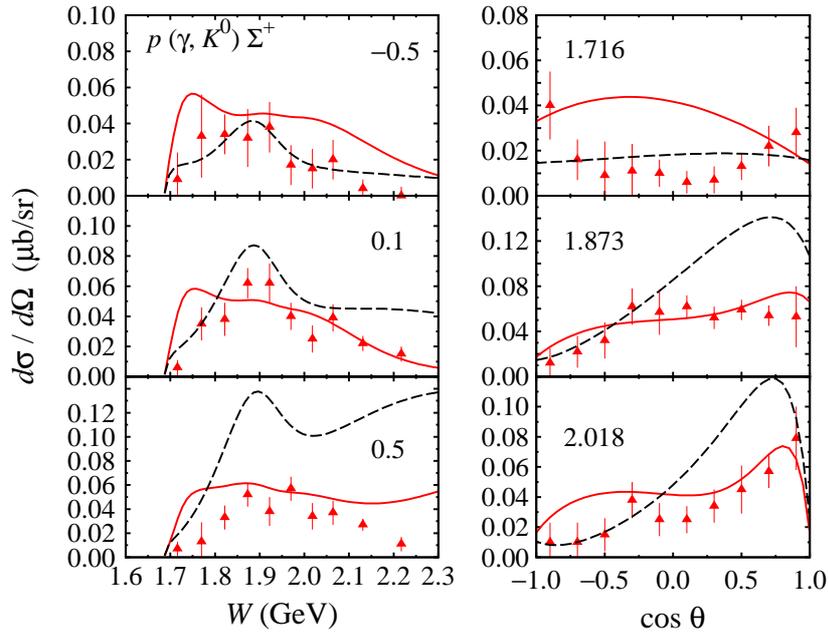}
\caption{Differential cross section for the $p(\gamma,K^0)\Sigma^+$ channel. Solid lines show the result of fitting to the new elementary reaction data only. Dashed lines display KAON-MAID result. Experimental data are taken from Ref.~\cite{lawall05}.}
\label{dk0sp_el} 
\end{figure*}

\begin{figure*}
\centering
  \includegraphics[width=0.80\textwidth]{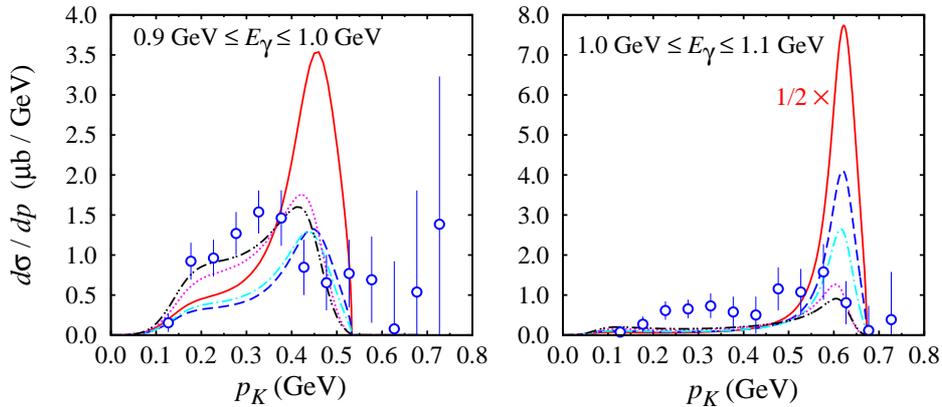}
\caption{Inclusive cross section of $d(\gamma,K^0)YN$ process. Solid lines show result of fitting to the data of elementary reaction data only. Other lines are obtained from fitting to the elementary as well as deuteron data by using a weighting factor $w$. Dashed line, dash-dotted line, dotted line, and dash-dot-dotted line are obtained by using $w=1$, 10, 100, and 1000, respectively. Experimental data are taken form Ref.~\cite{Tsu08}.}
\label{deuteron_new}
\end{figure*}

Fitting to the experimental data has been done in two ways. First we fit the model to some new data of elementary reactions only~\cite{lawall05}. Figure~\ref{dk0sp_el} shows the differential cross section for $K^0\Sigma^+$ channel after this fitting procedure. In this paper we only show this channel due to the lack of space. From this figure it seems to us that the new model provide an important improvement to the Kaon-Maid especially at higher energies and forward angles. The extracted parameters obtained from this fitting is then used to calculate the inclusive cross section of the $d(\gamma,K^0)\Lambda p$ process, as shown by the solid line in Fig.~\ref{deuteron_new}. Obviously, the model still overpredicts the experimental data.

Then, we fit the parameters in the model to all data, including the deuteron ones. In this case we introduce a weighting factor $w$ in the fitting procedure to emphasize the importance of the deuteron data. Explicitly, we use  
$\chi^2 = \chi^2_{\rm{elementary}}+w\times\chi^2_{\rm{deuteron}}$, 
in order to get a better agreement between our fit and deuteron data. The results are shown in Fig.~\ref{deuteron_new}, where it is clear that the weighting factor leads to a significant improvement of the model.

\begin{figure*}
\centering
  \includegraphics[width=1.\textwidth]{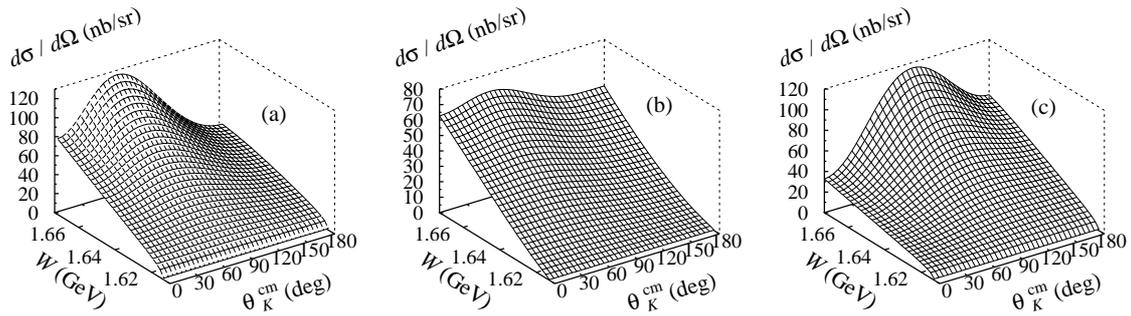}
\caption{Three dimensional plot of differential cross section for $n(\gamma,K^0)\Lambda$ channel. Left panel is the prediction of the present work, middle panel shows the model that fits to the elementary data near threshold \cite{Mar11}, and right panel is the prediction of KAON-MAID \cite{MaB00}.}
\label{dif_3db} 
\end{figure*}

Figure~\ref{dif_3db} shows the calculated differential cross section for the $n(\gamma,K^0)\Lambda$ channel as functions of the total c.m. energy and kaon scattering angle. The left panel in this figure is obtained from fitting to all data, the middle panel is the prediction by using the parameter values from fitting to the data of elementary reaction only, and the right panel is KAON-MAID result. Obviously, fittings to the new data do not indicate a backward peaking features in the angular distribution of $n(\gamma,K^0)\Lambda$ channel. A more detailed result of our calculation will be published soon \cite{salam_submitted}.

\section{Conclusion}
\label{conclusion}

We have investigated kaon photoproduction of the nucleon and deuteron. The parameters obtained from fitting to the new elementary reaction data only still leads to an overprediction of the inclusive reaction cross section on deuteron. More experimental data of kaon photoproduction on the deuteron are needed in order to get a better understanding of kaon photoproduction process as well as to avoid the weighting factor used in the fitting procedure. There is no indication of the backward peaking feature in the angular distribution of $n(\gamma,K^0)\Lambda$ channel obtained in the present work. 

\begin{acknowledgements}
This work has been supported by University of Indonesia under the Cluster Research Grant Scheme. 
\end{acknowledgements}


\end{document}